\documentclass[papersize,journal]{IEEEtran}
\usepackage{amsmath,amsfonts}
\usepackage{algorithmic}
\usepackage{algorithm}
\usepackage{array}
\usepackage[caption=false,font=normalsize,labelfont=sf,textfont=sf]{subfig}
\usepackage{textcomp}
\usepackage{stfloats}
\usepackage{url}
\usepackage{verbatim}
\usepackage{graphicx}
\usepackage{cite}
\usepackage[hidelinks,
linkcolor=black,
anchorcolor=black,
citecolor=black]{hyperref}

\begin{document}
\title{GI-Free Pilot-Aided Channel Estimation for Affine Frequency Division Multiplexing Systems}

\author{Yu Zhou, Haoran Yin, Nanhao Zhou, Yanqun Tang, Xiaoying Zhang, and Weijie Yuan, ~\IEEEmembership{Member,~IEEE}
\vspace{-1.5em}
\thanks{This work was supported by the Guangdong Natural Science Foundation under Grant 2019A1515011622, and the Hunan Natural Science Foundation under Grant 2022JJ30047. \textit{(Corresponding authors: Yanqun Tang;
Xiaoying Zhang.)}}
\thanks{Yu Zhou, Haoran Yin, Nanhao Zhou, and Yanqun Tang are with the School of Electronics and Communication Engineering, Sun Yat-sen University, Shenzhen 518107, China (email: 
\{zhouy633; yinhr6; zhounh3\}@mail2.sysu.edu.cn;
tangyq8@mail.sysu.edu.cn).}
\thanks{Xiaoying Zhang is with the College of Electronic Science and
Technology, National University of Defense Technology, Changsha 410073,
China (e-mail: zhangxiaoying@nudt.edu.cn).}

\thanks{Weijie Yuan is with the Department of Electronic and Electrical
Engineering, Southern University of Science and Technology,
Shenzhen 518055, China (e-mail: yuanwj@sustech.edu.cn).}
}



\maketitle

\begin{abstract}
The recently developed affine frequency division multiplexing (AFDM) can achieve full diversity in doubly selective channels, providing a comprehensive sparse representation of the delay-Doppler domain channel. 
Thus, accurate channel estimation is feasible by using just one pilot symbol. However, traditional AFDM channel estimation schemes necessitate the use of guard intervals (GI) to mitigate data-pilot interference, leading to spectral efficiency degradation.
In this paper, we propose a GI-free pilot-aided channel estimation algorithm for AFDM systems, which improves spectral efficiency significantly. 
To mitigate the interference between the pilot and data symbols caused by the absence of GI, we perform joint interference cancellation, channel estimation, and signal detection iterately.
Simulation results show that the bit error rate (BER) performance of the proposed method can approach
the ideal case with perfect channel estimation.
\end{abstract}

\begin{IEEEkeywords}
AFDM, doubly selective channels, channel estimation, data-pilot interference, interference cancellation.
\end{IEEEkeywords}

\section{Introduction}
\IEEEPARstart{O}{ne} of the most important scenarios of the next-generation cellular systems are extremely high mobility communications such as high-speed train, vehicle-to-vehicle, and vehicle-to-infrastructure. Orthogonal frequency division multiplexing (OFDM) is a widely used modulation technique in existing cellular systems. 
It demonstrates outstanding performance in time-invariant frequency selective channels. 
Nevertheless, in high-mobility environments characterized by significant Doppler frequency shifts, the orthogonality among OFDM subcarriers is compromised, which leads to serious performance degradation \cite{ref1}. Therefore, new modulation schemes are urgently called for supporting future mobile systems with high mobility.

Orthogonal time frequency space (OTFS) modulation is a
new modulation scheme adopting delay-Doppler (DD) domain for multiplexing the information symbols
instead of time-frequency (TF) domain as in conventional modulation schemes \cite{ref2}.
By spreading information symbols over the whole TF resources, the OTFS modulation shows superior performance over the classic OFDM modulation in high-mobility environment \cite{ref4}.
In \cite{ref5}, OTFS channel estimation based on one pilot symbol was proposed while OTFS suffers from excessive pilot overhead due to its two-dimensional (2D) structure as each pilot symbol should be separated from the data symbols.

Recently, Bemani \textit{et al}. developed a novel multicarrier modulation scheme called affine frequency division multiplexing (AFDM), which has 
shown almost the same bit error rate (BER) performance with OTFS \cite{ref6}.
However, AFDM can achieve the optimal diversity order of the doubly selective channels unlike OTFS \cite{ref7}.
AFDM is based on discrete affine Fourier transform (DAFT),  which is a generalization of discrete Fourier transform (DFT) and characterized by two parameters that can be adapted based on the Doppler spread of the doubly selective channels.
AFDM can attain a complete sparse representation of the DD channel, ensuring that all paths are distinctly separated. 
Consequently, each transmitted symbol in AFDM undergoes all the propagation paths.

The channel state information (CSI) is typically unknown, underscoring the essential role of channel estimation in facilitating precise symbol detection. 
Therefore, some investigations of channel estimation have been conducted in the AFDM literature. 
In \cite{ref7}, embedded pilot-aided (EPA) channel estimation based on one pilot symbol was proposed for AFDM by exploring its inherent full DD sparse
representation. 
Then, a low-complexity EPA diagonal reconstruction channel estimation scheme is proposed \cite{ref8}, which calculates the AFDM effective channel matrix directly without estimating the channel parameters.
For highly accurate estimation, \cite{ref7} and \cite{ref8} require the insertion of guard intervals (GI) to avoid interference between the pilot
and data symbols.
However, the adoption of GI will degrade the spectral efficiency since only fewer data symbols can be carried in one AFDM frame, especially for the low latency communication scenarios requiring small frame sizes.
Therefore, an AFDM channel estimation scheme with enhanced
spectral efficiency needs to be investigated.

In this paper, we present a novel GI-free pilot-aided channel estimation algorithm for AFDM systems. 
We propose a novel frame structure at the transmitter where no GI is required between the pilot and data symbols. 
In contrast to conventional approaches introducing GI to mitigate interference, our proposed method notably enhances spectral efficiency.
Differing from the frame structure delineated in \cite{ref9,ref10}, the proposed frame structure in this paper maintains a separation between the pilot and data symbols, i.e., there is no superimposition. 
This design enables the utilization of uncomplicated and practical detectors, such as the linear minimum mean square error (LMMSE) \footnote{In this paper, we use a simple and practical LMMSE detector to validate the robustness of the proposed GI-free pilot-aided channel estimation algorithm.}.
Due to the absence of the GI between the pilot and data symbols, the interference is unavoidable.
The inevitable interference between the pilot and data symbols is categorized into interference from data to pilot (ID2P) and interference from pilot to data (IP2D). 
The former affects the
accuracy of channel estimation, while the latter compromises the
precision of data detection.
To mitigate the two interference, 
we employ an iterative approach
involving interference cancellation, channel estimation, and
data detection to achieve accurate symbol detection.
We first eliminate the ID2P and optimize channel estimation under an
adjusted threshold. 
Subsequently, leveraging the refined channel estimation, we eliminate the IP2D to achieve precise data detection. 
By iterating the aforementioned two processes, a good channel estimation and signal detection performance can be obtained with only a few iterations.
The effectiveness of the algorithm has been verified through simulations. 
The proposed algorithm achieves a similar BER performance to the ideal case with perfect channel estimation while enhancing the spectral efficiency significantly.

\textit{Notations:}
The superscripts (·)$ ^T $ is the transpose operator, (·)$ ^* $ is the conjugate operator, and (·)$ ^H $ is the Hermitian transpose operator; 
$|\cdot|$ denotes the modulus of a complex number or the cardinality of a set;
$(\cdot)_N$ denotes the modulo operation with divisor $N$;
$\mathcal{C N}(m,v) $ denotes a complex Gaussian distribution with mean $m$ and variance $v$;
$ \textbf{I} $ denotes the identity metrics.

\section{Review of AFDM Systems}
\begin{figure}[!t]
\centering
\includegraphics[width=0.49\textwidth]{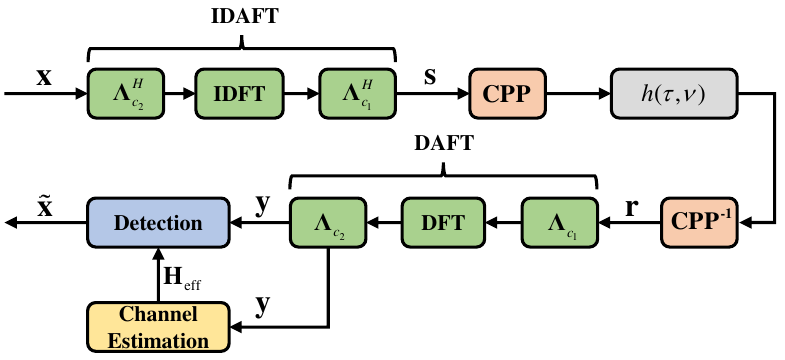}
\caption{The block diagram of AFDM systems.}
\label{fig_1}
	\vspace{-1.4em}
\end{figure}

In the following, we discuss the basic AFDM transceiver
structure based on \cite{ref7}. 
The block diagram of AFDM systems is shown in Fig. \ref{fig_1}.
Inverse DAFT (IDAFT) is firstly used to map data symbols into the time domain, while DAFT is performed at the receiver.
Let $\textbf{x} \in \mathbb{A}^{N \times 1}$ denote the vector of information symbols in the DAFT domain, where $ \mathbb{A} $ represents the constellation
set.
The modulated signal can be written as
\begin{equation}
\label{eq1}
s[n]=\frac{1}{\sqrt{N}} \sum_{m=0}^{N-1} x[m] e^{j 2 \pi\left(c_{1} n^{2}+\frac{1}{N} m n+c_{2} m^{2}\right)},
\end{equation}
where $c_1$ and $c_2$ are two AFDM parameters.
In matrix form, \eqref{eq1} becomes $\textbf{s} = {\bf{\Lambda }}_{{c_1}}^H{{\textbf{F}}^H}{\bf{\Lambda }}_{{c_2}}^H{\textbf{x}}$, where $\bf F$ is the DFT matrix, ${{\bf{\Lambda }}_{{c_1}}} = {\mathop{\rm diag}\nolimits} ({{e^{ - j2\pi {c_1}{n^2}}},n = 0,1, \ldots ,N - 1})$ and ${{\bf{\Lambda }}_{{c_2}}} = {\mathop{\rm diag}\nolimits} ({{e^{ - j2\pi {c_2}{m^2}}},m = 0,1, \ldots ,N - 1})$.
Due to different signal periodicity, a $L_{\rm cp}$-long chirp-periodic prefix (CPP) is used here, which plays the same role as the cyclic prefix (CP) in OFDM. 
$L_{\rm cp}$ is any integer greater than or equal to the value in samples of the maximum delay spread of the channel. The prefix is
\begin{equation}
s[n] = s[N + n]{e^{ - j2\pi {c_1}\left( {{N^2} + 2Nn} \right)}}, n =  - {L_{\rm cp}}, \cdots , - 1.
\end{equation}

The channel representation in the DD domain is given by
\begin{equation}
h(\tau ,\nu ) = \sum\limits_{i = 1}^P {{h_i}} \delta (\tau  - {\tau _i})\delta (\nu  - {\nu _i}),
\end{equation}
with $P$ the number of resolvable reflectors, $h_i$ the channel gain, $\tau_i$ the delay, and $\nu_i$ the Doppler shift of the $i$-th path, respectively.
Considering an AFDM symbol with a duration of $T$ and bandwidth of $B=N\Delta f$, where $\Delta f = 1/T$ denote the subcarrier spacing.
We define ${\tau _i} = l_i\frac{{1}}{{N\Delta f}}$ and ${\nu _i} =k_i \frac{{1}}{T}$, where $l_i$ and $k_i$ are the delay index and Doppler
index associated with the $i$-th path, respectively. We assume the dalay and Doppler indices are integer.

After transmission over the channel, the received samples are
\begin{equation}
\label{eq4}
r[n] = \sum\limits_{i = 1}^P {{h_i}} {e^{j2\pi \frac{{{k_i}\left( {n - {l_i}} \right)}}{N}}}s[(n - l_i)_N] + w[n],
\end{equation}
where $w[n]$ is the additive white Gaussian noise process with power spectral density (PSD) $N_0$.
We can write \eqref{eq4} in the matrix form as $\textbf{r} = \bf{H}\textbf{s} + \textbf{w}$, where $\bf{H}$ is the matrix representation of the time domain channel and $ \textbf{w} $ is the vector representation of $w[n]$.

At the receiver side, the DAFT domain output symbols are obtained by
\begin{equation}
\label{eq5}
	y[m]=\frac{1}{\sqrt{N}} \sum_{n=0}^{N-1} r[n] e^{-j 2 \pi\left(c_{1} n^{2}+\frac{1}{N} m n+c_{2} m^{2}\right)}.
\end{equation}
In matrix representation, \eqref{eq5} can be written as
\begin{equation}
\label{eq6}
\textbf{y} = {{\bf{\Lambda }}_{{c_2}}}{\bf{F}}{{\bf{\Lambda }}_{{c_1}}}\textbf{r} = {{\bf{\Lambda }}_{{c_2}}}{\bf{F}}{{\bf{\Lambda }}_{{c_1}}}{\bf{H\Lambda }}_{{c_1}}^H{{\bf{F}}^H}{\bf{\Lambda }}_{{c_2}}^H\textbf{x} + \tilde{\textbf{w}} = {{\bf{H}}_{{\rm{eff}}}}\textbf{x} + \tilde{\textbf{w}},
\end{equation}
where ${{\bf{H}}_{{\rm{eff}}}} = {{\bf{\Lambda }}_{{c_2}}}{\bf{F}}{{\bf{\Lambda }}_{{c_1}}}{\bf{H\Lambda }}_{{c_1}}^H{{\bf{F}}^H}{\bf{\Lambda }}_{{c_2}}^H$ and $\tilde{\textbf{w}} = {{\bf{\Lambda }}_{{c_2}}}{\bf{F}}{{\bf{\Lambda }}_{{c_1}}}{\textbf{w}}$.

The input-output relation of AFDM in \cite{ref7} can be represented as
\begin{equation}
\label{eq7}
y[m] = \sum\limits_{i = 1}^P {{h_i}} {e^{j\frac{{2\pi }}{N}\left( {N{c_1}l_i^2 - q_i{l_i} + N{c_2}\left( {{q_i^2} - {m^2}} \right)} \right)}}x[q_i] + \tilde w[m],
\end{equation}
where $q_i = {(m + {{\mathop{\rm loc}\nolimits} _i})_N}$ and ${{\mathop{\rm loc}\nolimits} _i} \buildrel \Delta \over = {({k_i} + 2N{c_1}{l_i})_N}$.
By tuning $c_1=\frac{{2{k_{\max }} + 1}}{{2N}}$ and setting $c_2$ to be an arbitrary irrational number or a rational number sufficiently smaller than $\frac{1}{2N}$ to achieve the full diversity in doubly selective channels.

\section{GI-Free Pilot-Aided Channel Estimation for AFDM Systems}
\subsection{Pilot Placement}
\begin{figure*}[!t]
\centering
\includegraphics[width=0.91\textwidth]{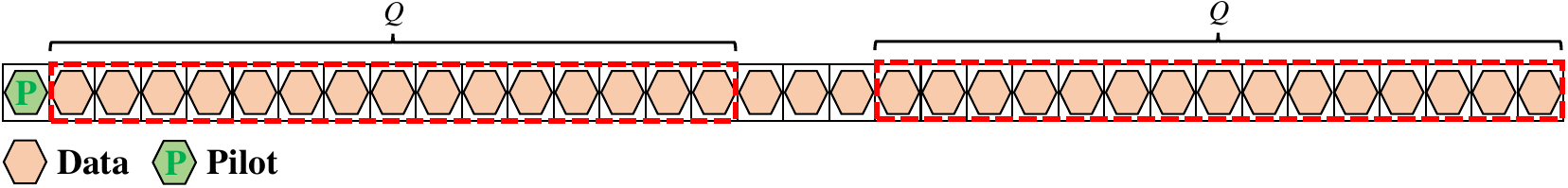}
\caption{Symbol arrangement of the proposed GI-free pilot-aided AFDM system at the transmitter.}
\label{fig_2}
	\vspace{-1.2em}
\end{figure*}

\begin{figure*}[!t]
\centering
\includegraphics[width=0.91\textwidth]{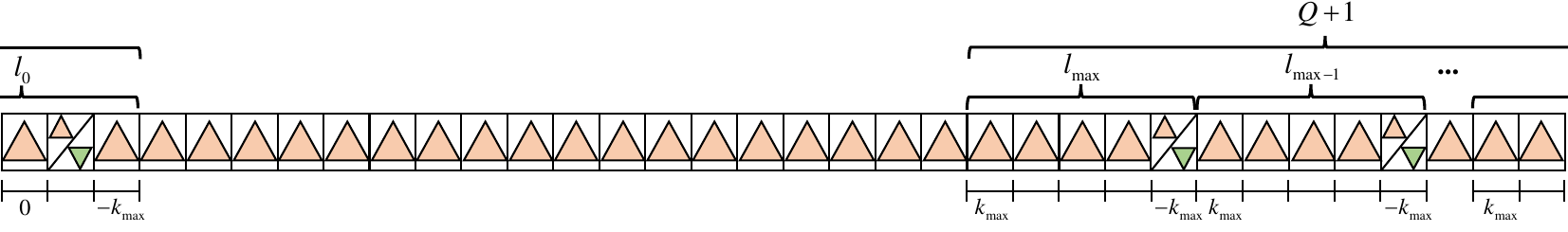}
\caption{Received frame structure of the proposed GI-free pilot-aided AFDM system at the receiver.}
\label{fig_3}
	\vspace{-1.2em}
\end{figure*}

For the classic approach \cite{ref7}, given the maximum delay index and Doppler index given by $l_{\rm max}$ and $k_{\rm max}$, channel estimation can be done through the following pilot placement
\begin{equation}
x[m]=\left\{\begin{array}{ll}
{x_{{\rm{pilot}}}}, & m=0 \\
0, & 1 \le m \le Q,N-Q \le m \le N-1 \\
{x_{{\rm{data}}}}, & Q+1 \le m \le N - Q - 1
\end{array}\right.,
\end{equation}
where
\begin{equation}
Q \buildrel \Delta \over = \left( {{l_{\max }} + 1} \right)\left( {2{k_{\max }} + 1} \right) - 1.
\end{equation}
We can see that a total number of $2Q+1$ grids are used for AFDM channel estimation. Therefore, the spectral efficiency can be defined as $\eta  = \frac{({N - 2Q - 1})\log\mathbb{|A|}}{N}{\rm{ bits/s/Hz}}$.
For a high-mobility scenario with a high Doppler shift, we need a large value of $Q$ to avoid the interference between the pilot and data symbols.
Meanwhile, in future wireless communications, there is a highly stringent requirement for low latency. In practical terms, this necessitates a relatively short AFDM frame duration, imposing a significant constraint on the size of the AFDM frame. 
In such scenarios, the spectral efficiency is substantially affected.
Hence, in the remainder of this section, we propose a GI-free pilot-aided channel estimation scheme in which only one pilot symbol is used for channel estimation.
The symbol arrangement is depicted in Fig. \ref{fig_2}, where the area bounded by the red dashed line denotes the GI adopted in the classic AFDM channel estimation \cite{ref7}.

As illustrated in Fig. \ref{fig_2}, we propose the GI-free pilot-aided scheme that utilizes the following pilot placement
\begin{equation}
x[m] = \left\{ \begin{array}{ll}
{x_{{\rm{pilot}}}}, & m=0 \\
{x_{{\rm{data}}}}, & 1 \le m \le N - 1
\end{array} \right..
\end{equation}
We denote the pilot symbol energy by $ E_p = |x_{\rm pilot}|^2 $ and the
average data symbol energy by $ E_{s}=\mathbb{E}\left\{|x_{\rm data}|^{2}\right\} $.
We can observe that only one pilot grid is occupied
for AFDM channel estimation.
However, due to the absence of the GI between the pilot and data symbols, the interference is unavoidable.
The received samples related to the
pilot, i.e., the $Q+1$ positions illustrated in Fig. \ref{fig_3}, can be rewritten as
\begin{equation}
\label{eq11}
y[m] = \mathcal{I}_{\rm D2P}[m] + \mathcal{I}_{\rm P2D}[m] + \tilde w[m],
\end{equation}
where $\mathcal{I}_{\rm D2P}[m]$ and $\mathcal{I}_{\rm P2D}[m]$ denote the interference introduced by the data symbols and pilot symbol, respectively. According to \eqref{eq7}, $\mathcal{I}_{\rm D2P}[m]$ and $\mathcal{I}_{\rm P2D}[m]$ are given by
\begin{equation}
\label{eq12}
\mathcal{I}_{\rm D2P}[m] = \sum\limits_{i=1}^P {{h_i}{e^{j\frac{{2\pi }}{N}(N{c_1}l_i^2 - {q_i}{l_i} + N{c_2}({q^2_i} - {m^2}))}}x[q_i]} ,q_i \ne 0
\end{equation}
and
\begin{equation}
\label{eq13}
\mathcal{I}_{\rm P2D}[m] = \sum\limits_{i=1}^P {{h_i}{e^{j\frac{{2\pi }}{N}(N{c_1}l_i^2 - {q_i}{l_i} + N{c_2}({q^2_i} - {m^2}))}}x[q_i]} ,q_i = 0.
\end{equation}
Thus \eqref{eq13} can be rewritten as
\begin{equation}
\label{eq14}
\mathcal{I}_{\rm P2D}[m] = \sum\limits_{i=1}^P {{h_i}{e^{j 2\pi ({c_1}l_i^2 - {c_2}{m^2})}}x_{\rm pilot}}.
\end{equation}
In Fig. \ref{fig_3}, the frame structure at the receiver is depicted, where green triangles denote the response to pilot at specific samples, and orange triangles represent the response to data at those samples. 
It is evident that the pilot can induce responses only at the $Q+1$ samples illustrated in Fig. \ref{fig_3}. 
Responses at other samples are exclusively generated by data and noise. 
In a channel with $P$ paths, the pilot induces responses at $P$ samples among these $Q+1$ samples. 
These $P$ samples represent the locations where interference is introduced by the pilot and data symbols. 
In the following section, we propose an iterative GI-free pilot-aided channel estimation algorithm
for AFDM communication systems.
	\vspace{-1.5em}

\subsection{The Proposed GI-Free Pilot-Aided Channel Estimation Scheme}
\begin{figure}[!t]
\centering
\includegraphics[width=0.476\textwidth]{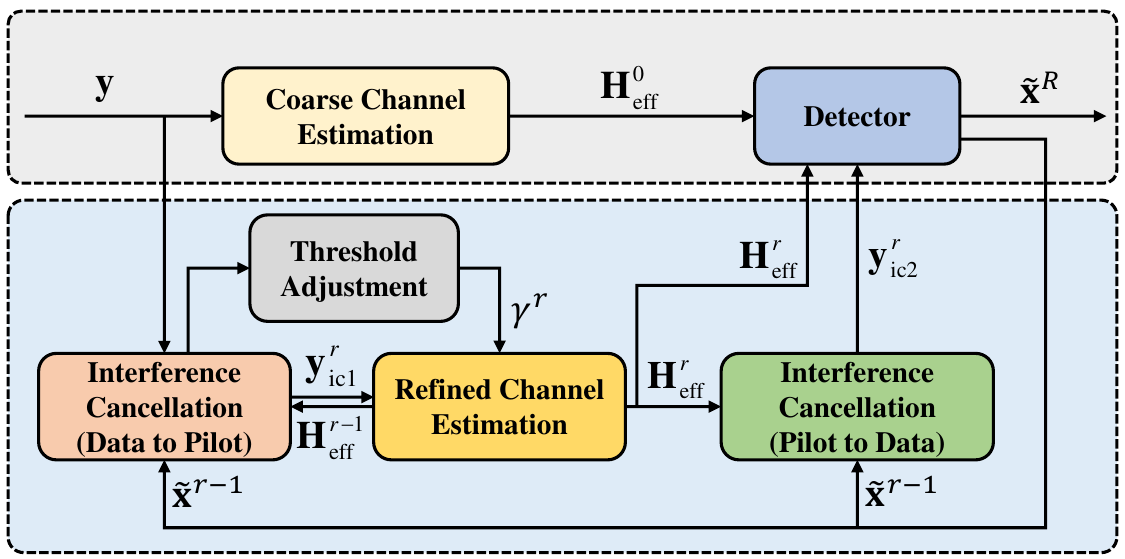}
\caption{Block diagram of the proposed GI-free pilot-aided AFDM channel estimation algorithm.}
\label{fig_4}
	\vspace{-1.2em}
\end{figure}

The block diagram of the proposed GI-free pilot-aided AFDM channel estimation scheme is given in Fig. \ref{fig_4}.
We classify interference between the pilot and data symbols into ID2P and IP2D. 
The former affects the
accuracy of channel estimation, while the latter impacts the
precision of data detection.
Consequently, we will address and eliminate these two types of interference separately to achieve accurate channel estimation and precise data detection.
Due to the existence of ID2P, the channel parameters can not be simply obtained using the threshold $\gamma  = 3\sqrt {N_0} $ used in \cite{ref7} for channel estimation.

Under a modified threshold $\gamma$ considering interference $\mathcal{I}_{\rm D2P}[m]$, we compare the received signal $y[m]$ with $\gamma$ to obtain a coarse estimation of the channel parameters. According to \eqref{eq12}, we can calculate the interference energy, expressed as
\begin{equation}
\label{eq15}
\begin{array}{l}
\mathbb{E}\left\{\left|\mathcal{I}_{\rm D2P}[m]\right|^{2}\right\}=\sum\limits_{i=1}^P \mathbb{E}\left\{\left|h_{i}\right|^{2}\right\} \mathbb{E}\left\{\left|x\left[q_i\right]\right|^{2}\right\}  + \\
\quad \sum\limits_{i\ne i^{\prime}}^P \mathbb{E}\left\{h_{i} h_{i^{\prime}}^{*}\right\} \mathbb{E}\left\{x\left[q_i\right] \cdot x^{*}\left[q_{i^{\prime}}\right]\right\}
e^\zeta
\end{array}
\end{equation}
with $ e^\zeta = e^{j\frac{{2\pi }}{N}(N{c_1}(l_i^2-l^2_{i^{\prime}}) - {q_i}{l_i} + {q_{i^{\prime}}}{l_{i^{\prime}}} + N{c_2}({q^2_i} - {q^2_{i^{\prime}}}))} $, provided that data symbols are independent from the channel coefficients. 
Assuming that the channel coefficients are independent for different paths $\mathbb{E}\left\{h_{i} h_{i^{\prime}}^{*}\right\}=0$, we can further simplify \eqref{eq15} as
\begin{equation}
\label{eq16}
\begin{array}{l}
\mathbb{E}\left\{\left|\mathcal{I}_{\rm D2P}[m]\right|^{2}\right\}=\sum\limits_{i=1}^P \mathbb{E}\left\{\left|h_{i}\right|^{2}\right\} E_s.
\end{array}
\end{equation}

In this paper, we assume that the channel gains obey the
exponential power delay profile. 
Note that the channel power
gains are in general normalized, yielding $\sum_{i=1}^P \mathbb{E}\left\{\left|h_{i}\right|^{2}\right\}=1$.
Therefore, the energy of the interference term is given by $\mathbb{E}\left\{\left|\mathcal{I}_{\rm D2P}[m]\right|^{2}\right\}=E_s$.
Consequently, we have the modified threshold for coarse channel estimation, given by
\begin{equation}
\gamma  = 3\sqrt {{N_0} + {E_s}}.
\end{equation}
If $|y[m]| \geq \gamma$, the delay index and Doppler shift index can be directly obtained as depicted in Fig. \ref{fig_3}, and the channel gain can be obtained as
\begin{equation}
\label{eq18}
\tilde{h}_i = \frac{y[m]}{e^{j2\pi\left( {{c_1}\tilde{l}_i^2 - {c_2}{m^2}} \right)}x_{\rm pilot}}.
\end{equation}
Iterate over all $Q+1$ samples related to the pilot to obtain estimates of the channel parameters, i.e., $\{ {{{\tilde h}_1},{{\tilde h}_2}, \cdots ,{{\tilde h}_{\tilde P}}} \}$, $\{ {{{\tilde l}_1},{{\tilde l}_2}, \cdots ,{{\tilde l}_{\tilde P}}}\}$, and $\{ {{{\tilde k}_1},{{\tilde k}_2}, \cdots ,{{\tilde k}_{\tilde P}}}\}$. 
We can utilize the obtained coarse channel parameters to derive the equivalent channel matrix $ {\bf{H}}_{\rm eff}^0 $.

After obtaining the channel estimates, we utilize an LMMSE detector for symbol detection to obtain the coarse data detection, given by
\begin{equation}
\tilde{\textbf{x}}^0 = {({{\bf{H}}_{\rm eff}^0}^H{\bf{H}}_{\rm eff}^0 + {N_0}{\bf{I}})^{ - 1}}{{\bf{H}}_{\rm eff}^0}^H \textbf{y}.
\end{equation}
Due to the impact of noise and detection performance, the cancellation is generally
imperfect. 
Therefore, we employ a scheme that iteratively performs interference cancellation, channel estimation, and data detection. 
Improved channel estimation is expected to refine data detection, contributing to a more accurate interference cancellation process crucial for executing threshold-based channel estimation.

For the $r$-th iteration, according to \eqref{eq12}, we are able to cancel out
ID2P based on the
estimated channel and the obtained symbols, given by
\begin{equation}
y^r_{\rm ic1}[m] = y[m]- \sum\limits_{i=1}^{{\tilde P}^{r-1}} {{\tilde{h}^{r-1}_i}{e^{\varsigma}}{\tilde x}^{r-1}[q_i]}
\end{equation}
with $ e^{\varsigma}={e^{j\frac{{2\pi }}{N}(N{c_1}(\tilde{l}^{r-1}_i)^2 - {q_i}{\tilde{l}^{r-1}_i} + N{c_2}({q^2_i} - {m^2}))}}$ and $q_i = {(m + ({\tilde{k}^{r-1}_i} + 2N{c_1}{\tilde{l}^{r-1}_i})_N)_N}$.
Eliminating ID2P refers to the removal of the orange triangles at the samples where the orange triangles overlap with the green triangles, as depicted in Fig. \ref{fig_3}.
If the interference is perfectly cancelled, then the residual term $y^r_{\rm ic1}[m]$
contains only pilot information and the noise.

In general, the number of channel paths is unknown to the receiver. 
As a compromise approach, we consider a relatively
large number $P'$ to avoid the misdetection of the channel paths with small gains.
Assuming that we have identified ${\tilde P}^{r-1}$ paths out of total $P'$ paths in the $ (r-1) $-th iteration, after eliminating ID2P, there are still interference corresponding to the
rest $P'-{\tilde P}^{r-1}$ paths. 
The energy for the residual ID2P
can be roughly obtained as $\frac{P'-{\tilde P}^{r-1}}{P'} E_s$. 
Consequently, the new
threshold for refined channel estimation at the $r$-th iteration is given by
\begin{equation}
{\gamma^r} = 3\sqrt {{N_0} + \frac{(P' - {\tilde P}^{r-1})}{P'}{E_s}}.
\end{equation}
Thus we can compare $y^r_{\rm ic1}[m]$ with ${\gamma^r}$ to obtain a refined channel estimation, and \eqref{eq18} becomes
\begin{equation}
\tilde{h}^r_i = \frac{y^r_{\rm ic1}[m]}{e^{j2\pi\left( {{c_1}({\tilde{l}^r}_i)^2 - {c_2}{m^2}} \right)}x_{\rm pilot}}.
\end{equation}

After obtaining the refined channel parameters, we can eliminate IP2D by
\begin{equation}
y^r_{\rm ic2}[m] = y[m] - \sum\limits_{i=1}^{{\tilde P}^{r}} {{\tilde{h}^r_i}{e^{j 2\pi ({c_1}({\tilde{l}^r}_i)^2 - {c_2}{m^2})}}x_{\rm pilot}}.
\end{equation}
Eliminating IP2D involves removing the green triangles at the samples where the orange triangles overlap with the green triangles, as shown in Fig. \ref{fig_3}.
We utilize the refined channel parameters to derive the equivalent channel matrix $ {\bf{H}}_{\rm eff}^r $. Then we can achieve accurate data detection, given by
\begin{equation}
\tilde{\textbf{x}}^r = {({{\bf{H}}_{\rm eff}^r}^H{\bf{H}}_{\rm eff}^r + {N_0}{\bf{I}})^{ - 1}}{{\bf{H}}_{\rm eff}^r}^H \textbf{y}^r_{\rm ic2},
\end{equation}
where $\textbf{y}^r_{\rm ic2}$ represents the vector representation of $y^r_{\rm ic2}[m]$.
After running $R$
iterations, the channel estimates as well as the detected data
symbols are obtained. 
In general, the iterative process can be
terminated if the performance gain by running more iterations
is marginal.

Since only one pilot symbol is used for channel estimation, it can be calculated that the spectral efficiency of
the proposed algorithm is $\eta  = \frac{({N - 1})\log\mathbb{|A|}}{N}{\rm{ bits/s/Hz}}$.
Compared to the classic channel estimation \cite{ref7}, the spectral efficiency has been improved by $\frac{N-1}{N-2Q-1} \times 100\% $, which is
significant when the size of the transmitted AFDM frame is
relatively small.
	\vspace{-1em}

\section{Simulation Results}
\begin{figure}[!t]
\centering
\includegraphics[width=0.47\textwidth]{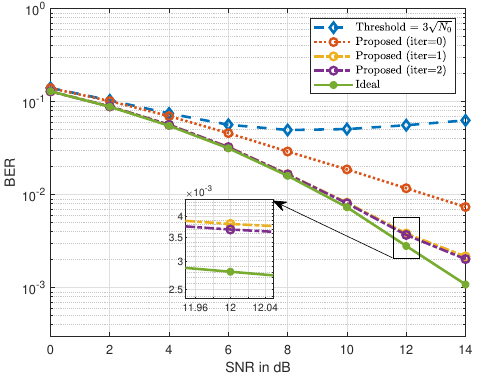}
\caption{BER performance versus SNR.}
\label{fig_5}
	\vspace{-1.8em}
\end{figure}
In this section, we provide simulation results to assess the
performance of the proposed algorithm.
We consider an AFDM frame with $ N = 512 $.
The carrier frequency is $4 \rm GHz$ and the subcarrier spacing is $1 \rm kHz$.
The maximum moving speed in the scenario is $1080 \rm km/h$, yielding a maximum
Doppler index with $k_{\rm max}=4$, and the maximum delay index
is $l_{\rm max} = 10$. 
The number of channel paths is $P = 3$, the associated delay index for each channel path is
randomly chosen from $[0,l_{\rm max}]$, and Jakes Doppler spectrum is considered, i.e., the
Doppler indices are varying and the Doppler index of each path
is generated using $k_i=k_{\rm max}\cos \theta_i$, where $\theta_i$
is uniformly distributed over $ [-\pi,\pi] $.
The channel gain is generated according the distribution $\mathcal{C N}(0,\frac{1}{P}) $.
The signal-to-noise ratio (SNR) for the transmission is defined as $ {\rm SNR}=\frac{E_s}{N_0} $.
We assume that the SNR of pilot is $ {{\rm SNR}_p}=\frac{E_p}{N_0}= 45 \rm dB $.
Binary phase shift keying (BPSK)
modulation is used for the bit mapping.
Finally, the maximum number of iterations is $R = 2$, and $P'=4$.

In Fig. \ref{fig_5}, we illustrate the BER performance based on the LMMSE detector.
We can observe that, due to the absence of GI between pilot and data symbols in the proposed algorithm, the BER performance using the classical channel estimation threshold is the poorest (Threshold = $3\sqrt{N_0} $). 
This also indicates that continuing to apply traditional thresholds is impractical.
Consequently, the threshold needs to consider the energy of ID2P, as derived earlier.
Under the adjusted threshold, it is evident that the algorithm's BER performance without iteration, i.e., directly using coarse channel estimation for data symbols detection, is significantly better than the performance without threshold adjustment. 
However, it remains notably inferior compared to the ideal case, where the channel parameters are perfectly known to
the detector. 
Due to the interference between the pilot and data symbols, the coarse channel estimation
by regarding all interference as noise suffers from significant performance loss.
Moreover, we can see that a single iteration yields an approximately $4 \rm dB$ gain compared to non-iterative approaches, and at low SNR, the BER performance after one iteration closely approaches the ideal case. 
Since we use a fixed pilot energy,
with the increase of the SNR, the ID2P becomes more severe, leading to an increase in the BER of the proposed algorithm.
The improvement in BER performance from one to two iterations is marginal, primarily due to the utilization of the LMMSE detector with limited diversity extraction capabilities. The potential for iterative enhancement could be further realized with the adoption of a more sophisticated detector with superior performance.
Our proposed algorithm achieves excellent BER performance with very few iterations, indicating that it does not incur significant additional computational overhead.
Additionally, our algorithm can achieve a spectral
efficiency of $ \eta = 0.998 \rm bits/s/Hz $ while the classic approach has $ \eta = 0.615 \rm bits/s/Hz $.
We can see that the GI-free channel estimation scheme achieves an improvement of $ 62.2\% $ of the spectral efficiency. 
This improvement becomes more attractive when supporting multiple users where more guard space is required
for avoiding inter-user interference.
As for the peak-to-average-power ratio (PAPR), with the symbol power of $ 10 \rm dB $ and the pilot power of $ 45 \rm dB $, the PAPR for the proposed symbol placement scheme is $4.704{\rm{dB}}$ compared to $4.007{\rm{dB}}$ for the conventional scheme with GI. Nevertheless, the spectral efficiency can be increased $ 62.2\% $, at the cost of only $17.4\% $ PAPR increase in dB.
	\vspace{-0.8em}

\section{Conclusion}
This paper proposed a GI-free pilot-aided channel estimation
algorithm for AFDM systems. 
In comparison with classical AFDM channel estimation approaches, we no longer incorporate GI between the pilot and data symbols, which can notably enhance spectral efficiency, especially in low-latency communication scenarios where the frame size cannot be too large.
The inevitable interference between the pilot and data symbols is categorized into interference from data to pilot and interference from pilot to data. 
The former affects the accuracy of channel estimation, while the latter impacts the precision of data detection.
We employ an iterative approach involving interference cancellation, channel estimation, and data detection to achieve accurate symbol detection. Simulation results demonstrate that the performance of the proposed method can approach the ideal case with perfect channel estimation while improving spectral efficiency significantly.

\end{document}